\documentclass{Interspeech2024}
\usepackage{float}
\usepackage{caption}

\interspeechcameraready 
\title{Exploring bat song syllable representations in self-supervised audio encoders}

\name[affiliation={1}]{Marianne}{de Heer Kloots}
\name[affiliation={2}]{Mirjam}{Knörnschild}

\address{
  $^1$Institute for Logic, Language and Computation, University of Amsterdam; The Netherlands
  $^2$Museum für Naturkunde, Leibniz Institute for Evolution and Biodiversity Science; Germany}
\email{m.l.s.deheerkloots@uva.nl, mirjam.knoernschild@mfn.berlin}


\begin{document}

\maketitle

\begin{abstract}
How well can deep learning models trained on human-generated sounds distinguish between another species' vocalization types? We analyze the encoding of bat song syllables in several self-supervised audio encoders, and find that models pre-trained on human speech generate the most distinctive representations of different syllable types. These findings form first steps towards the application of cross-species transfer learning in bat bioacoustics, as well as an improved understanding of out-of-distribution signal processing in audio encoder models.
\end{abstract}

\section{Introduction}
Many researchers in bioacoustics would benefit from robust and accurate feature spaces that can handle graded vocalizations in real-world field recordings, for example for the purpose of automatic classification. In the domain of human speech and sound processing, much recent progress is driven by so-called self-supervised audio encoder models \cite{baevskiWav2vecFrameworkSelfSupervised2020, hsuHuBERTSelfSupervisedSpeech2021}, which learn rich representations of acoustic signals through a masked audio segment prediction task on unlabelled data. Training such models from scratch for non-human species is currently still infeasible, due to the limited size of most bioacoustic datasets \cite{manriquezDeepLearningSmall2024, stowellComputationalBioacousticsDeep2022}. However, existing pre-trained models still offer promising opportunities through their use in \emph{cross-species transfer learning}, providing a new tool to explore divergences and commonalities between species \cite{cauzinilleSpeechPrimateVocalizations2024}. Here, we explore how a variety of self-supervised audio models trained on human and non-human generated sounds encode bat song syllable types in field recordings of one species' territorial song.

\section{Data}
We use a dataset of 20 territorial songs produced by males of the Greater Sac-Winged Bat (\emph{Saccopteryx bilineata}), recorded in Costa Rica using an ultrasonic microphone (Avisoft USG 116Hme with condenser microphone CM16; frequency range 1–200 kHz). These multisyllabic vocalizations are acquired by imitation from tutor males during ontogeny \cite{knornschildVocalProductionLearning2014a} and 
%used to defend roosting territories and attract females, while also 
encode personal information about the singer such as individual identity, group affiliation and regional origin \cite{knornschildBatSongsAcoustic2017}. Territorial songs are composed of up to six different syllable types \cite{behrTerritorialSongsIndicate2006}, five of which are present in our dataset and manually labelled for analyses (420 syllables in total; including 135, 97, 92, 9, and 87 instances of syllable types A, B, C, D, and E, respectively).

\section{Analyses}

\subsection{Data pre-processing}
Several pre-processing steps were performed before feeding our dataset of \emph{S. bilineata} territorial songs through the pre-trained audio encoder models. For denoising, we used the noise reduction algorithm implemented in the software Avisoft SASLab Pro, which automatically recognizes syllables and removes noise below a user-defined threshold in the frequency domain. Depending on the noise floor of each recording, threshold levels were between -60 to -75 dB. Detected noise was reduced by 90dB. We further applied a high-pass filter of 10 kHz.

Vocalizations of the recorded \emph{S. bilineata} population have a species mean fundamental frequency (F0) around 15.5 kHz (SD: 2 kHz), but also contain much energy above 20 kHz. Such higher frequencies are mostly inaudible to humans and outside the training distribution of the pre-trained audio encoders studied here. After denoising, we therefore move the songs into the human auditory range by slowing down all recordings in our dataset by a factor of 8. In the slowed down recordings, mean syllable duration is 235 ms (SD: 135 ms) and most energy is contained within the 1-8 kHz frequency band for all syllable types (F0 mean: 2.3 kHz, SD: 900 Hz). Finally, we downsample all recordings to 16 kHz, as required for processing by the pre-trained audio encoders.

\subsection{Feature extraction}
Our set of four self-supervised models comprises two different architectures and three different sets of pre-training data (see \autoref{tab:models}). The AVES model is an audio representation model developed for encoding animal vocalizations; we here use the AVES-\texttt{bio}-base configuration pre-trained on a large set of animal sounds from various species. We also include another HuBERT-based model trained exclusively on human speech (Librispeech audiobooks \cite{panayotovLibrispeechASRCorpus2015}), as well as a Wav2Vec2.0 model trained on the same data, and a second Wav2Vec2.0 model trained exclusively on music (from the Free Music Archive \cite{defferrardFMADatasetMusic2017}). Each model consists of a CNN-based waveform encoder followed by 12 Transformer layers, ultimately generating 768-dimensional feature sequences at a frame rate of 20 ms. 

\begin{table}[h]
  \caption{Self-supervised audio models included in our analyses}
  \label{tab:models}
  \centering
  \begin{tabular}{lll}
    \toprule
    \textbf{Name}     & \textbf{Architecture} & \textbf{Training data}       \\
    \midrule
    AVES \cite{hagiwaraAVESAnimalVocalization2023}             & HuBERT       & 360h, animals \\
    HuBERT (speech) \cite{hsuHuBERTSelfSupervisedSpeech2021}   & HuBERT       & 960h, speech \\
    Wav2Vec2 (speech) \cite{baevskiWav2vecFrameworkSelfSupervised2020} & Wav2Vec2.0   & 960h, speech \\
    Wav2Vec2 (music) \cite{orhanAlgebraicStructuresEmerge2024} & Wav2Vec2.0   & 900h, music  \\  
    \bottomrule
  \end{tabular}
\end{table}

\begin{minipage}{\textwidth}
\hspace{-1.9em}
\makebox[\textwidth]{
  \includegraphics[width=\textwidth]{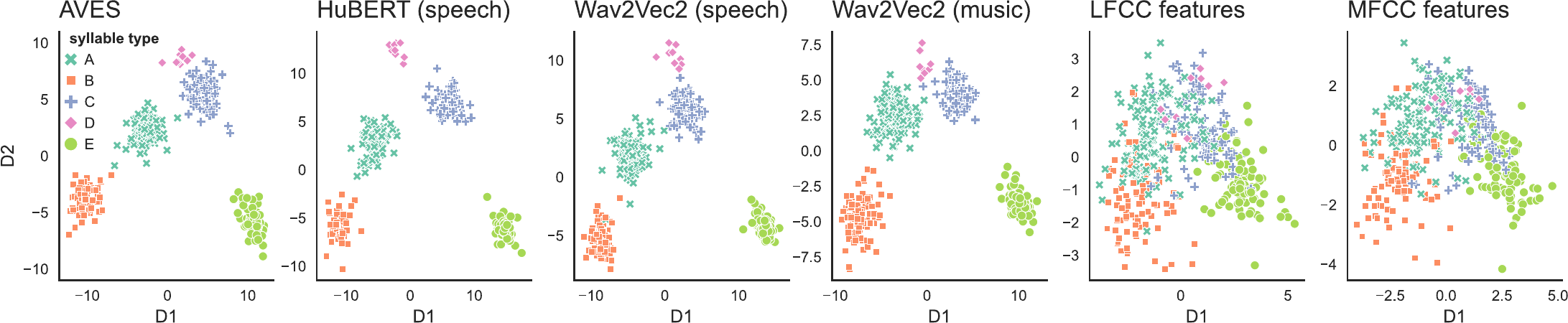}}
\captionof{figure}{Syllable projections along the two most discriminative directions in each LDA-transformed feature space.}\label{fig:lda-viz}
\end{minipage}

\begin{figure}[H]
\vspace{1em}
  \centering
  \includegraphics[width=\linewidth]{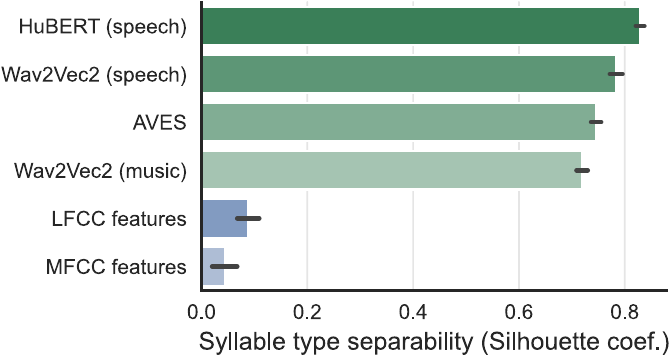}
  \caption{Separability between syllable type clusters is highest in the self-supervised models trained on human speech (error bars show 95\% confidence intervals).}
  \label{fig:separability}
\end{figure}

\noindent To create syllable representations using each of the self-supervised audio encoders, we pass a full song as input through the model, and extract frame representations from its final Transformer layer. We then average over all frames within each syllable, resulting in one 768-dimensional feature embedding for every syllable. Mean-pooling across time might seem overly simplistic for capturing distinct temporal dynamics between syllable types (e.g. upsweeps vs. downsweeps). However, similar mean-pooled Transformer-based embeddings have been shown to successfully capture information across several timescales in human speech processing (for example on the phoneme- \cite{pasadLayerWiseAnalysisSelfSupervised2021, klootsHumanlikeLinguisticBiases2024} and word-level \cite{pasadWhatSelfSupervisedSpeech2024a}), and perform well on bioacoustic transfer learning tasks \cite{hagiwaraAVESAnimalVocalization2023, cauzinilleInvestigatingSelfSupervised2024}.

Finally, we include two simpler feature sets of 13-dimensional linear- and Mel-frequency cepstral coefficients for comparison, each computed with a 400 sample FFT window.

\subsection{Separability analyses}

We aim to assess how distinctively \emph{S. bilineata} territorial song syllables are encoded in each feature space. For this purpose, we first project each set of syllable features into its 4 most discriminative directions using Linear Discriminant Analysis (LDA). \autoref{fig:lda-viz} visualizes every syllable's location along the first two directions of each projected feature space. This reveals that the self-supervised audio encoder models encode each of the 5 syllable types into distinguishable subspaces, which are linearly decodable from their final layer representations. In contrast, the LFCC and MFCC features show much more entanglement between syllable types. 

To more precisely quantify the separability between different syllable types in each feature space, we compute silhouette coefficients for each syllable type cluster based on Mahalanobis distances between samples. 

\newpage
\vspace*{14.5em} \noindent
The silhouette coefficient for each sample is defined as $(b - a) / \text{max}(a, b)$, where $a$ is the mean distance to all other points in the same cluster, and $b$ is the mean distance to all other points in the next nearest cluster. The mean silhouette coefficients per LDA-projected feature space are visualized in \autoref{fig:separability}. This shows that syllable type separability is highest in the two self-supervised models trained on human speech, followed by the model trained on animal vocalizations, and finally the model trained on music.

\section{Discussion \& Conclusions}
We find that the syllable types in our territorial song recordings, when slowed down to the human hearing range, are distinctively encoded by self-supervised audio encoders. Representations learned by such models thus encode useful features for \emph{S. bilineata} syllable identification, even when only pre-trained on sounds generated by other species. 

Interestingly, syllable types are most separable in the models pre-trained on human speech. This indicates that rich representations optimized for a single species' vocal repertoire might form a more promising basis for cross-species transfer learning than those optimized to encode a large variety of species, or non-vocal sound sources like musical instruments. However, the animal vocalization model included in our current comparison set was pre-trained on a substantially smaller amount of audio than the speech and music models (\autoref{tab:models}). A comparison against models pre-trained on fewer hours of speech would be needed to determine whether training dataset size could explain the difference between models trained on human speech vs. multiple species. Between the speech-trained models, the HuBERT architecture showed a slight syllable separability advantage compared to the Wav2Vec2 architecture. This could be due to the clustering objective that is part of the HuBERT training procedure \cite{hsuHuBERTSelfSupervisedSpeech2021}, potentially driving the model's internal representations towards generally more separable subspaces.

Our current findings indicate that self-supervised audio encoders pre-trained on human speech generate useful representations for distinguishing between \emph{S. bilineata} song syllable types. However, territorial songs in this species are known to also encode singer identity, and several other features \cite{knornschildBatSongsAcoustic2017} — models might differ in which features they most prominently encode. Representations from self-supervised models can be optimized by supervised fine-tuning to encode the most relevant features for specific classification and detection tasks.
In future work, we aim to further investigate what interpretable features contribute to the distinctive syllable type representations across each of the audio encoders' internal layers, and test the applicability of our approach to other tasks in bat bioacoustics, such as syllable detection, species and dialect identification. 

\clearpage

\section{Acknowledgements}
We thank Pierre Orhan and co-authors for sharing the weights of their music-trained Wav2Vec2 model with us, and two anonymous reviewers for very helpful comments.
\bibliographystyle{IEEEtran}
\bibliography{mybib}

\end{document}